\documentclass[twocolumn,
preprintnumbers,amsmath,amssymb]{revtex4}

\usepackage{graphicx}
\usepackage{dcolumn}
\usepackage{bm}

\begin{document}


\title{Topological Network Entanglement as \\
Order Parameter for the Emergence of Geometry}

\author{M. Cristina Diamantini}
\email{cristina.diamantini@pg.infn.it}
\affiliation{%
NiPS Laboratory, INFN and Dipartimento di Fisica, University of Perugia, via A. Pascoli, I-06100 Perugia, Italy
}%

\author{Carlo A. Trugenberger}
\email{ca.trugenberger@bluewin.ch}
\affiliation{%
SwissScientific, chemin Diodati 10, CH-1223 Cologny, Switzerland
}%

\date{\today}

\begin{abstract}
We show that, in discrete models of quantum gravity, emergent geometric space can be viewed as the entanglement pattern in a mixed quantum state of the ``universe", characterized by a universal topological network entanglement. As a concrete example we analyze the recently proposed model in which geometry emerges due to the condensation of 4-cycles in random regular bipartite graphs, driven by the combinatorial Ollivier-Ricci curvature. Using this model we show that the emergence of geometric order decreases the entanglement entropy of random configurations. The lowest geometric entanglement entropy is realized in four dimensions.

\end{abstract}
\maketitle

In recent years it has emerged that geometry and quantum entanglement are intimately connected \cite{rams1, rams2, rams3}. In a nutshell, the basic idea is to consider classical space-time as the entanglement in the quantum superposition of two disconnected copies of quantum states. This idea can be formalized via the AdS/CFT duality, since the machinery of CFTs permits explicit calculations and leads to expected results about black holes \cite{mald}. 

On the other side, one of us has recently proposed the idea that space-time is an emergent property of an ordered phase of fundamental constituents that are just information bits \cite{tru1}. Random bits are first assembled into random graphs \cite{graphrev}. Quantum gravity is then defined non-perturbatively by an ultraviolet (UV) quantum critical point corresponding to the condensation of short graph cycles and driven by the combinatorial version of Ricci curvature derived by Ollivier \cite{olli1}. In this quantum phase transition random graphs turn into geometrical graphs \cite{geom} and geometric space and the standard Einstein-Hilbert action emerge in the continuum limit.  

To test this idea a simplified ``toy" model was proposed \cite{tru2}, which consists in the restriction of the configuration space to random regular bipartite graphs, for which the Ollivier-Ricci curvature becomes analytically tractable \cite{olli2}. Monte Carlo simulations in this model clearly indicate the presence of two regimes for strong and weak gravity, corresponding to random regular bipartite graphs and geometric, locally $\mathbb{Z}^d$ lattices, respectively. Present data are, however not yet sufficient to definitively conclude that the crossover region for finite $N$ develops into a second-order phase transition when $N\to \infty$, although this picture looks very suggestive. Note also that the general idea is strongly supported by first results for generic geometric graphs \cite{krioukov}. 

In this note we show that these two approaches are very closely related. The crucial idea we use to this end is the realization that every classical graph, and thus all discrete models of quantum gravity, can also be viewed as the entanglement pattern in the quantum superposition of two disconnected quantum states \cite{zan}, the very mechanism originally proposed in the framework of the AdS/CFT duality. Very interestingly, as in topologically ordered states of matter \cite{top}, the von Neumann entanglement entropy contains a universal topological entanglement term \cite{kit} which characterizes the long distance quantum order associated with the graph. This was called the topological network entanglement in \cite{zan}.

We use the simplified model in \cite{tru2} to compute analytically the Renyi entanglement entropies and numerically the topological network entanglement. The Renyi entanglement entropies characterize the geometric order of the graphs via their dependence on the relative number of cycles of a given length. The topological network entanglement is higher in the geometric regime of the graphs than in their random phase. Its value in the geometric phase has a local maximum (corresponding to a local minimum in entanglement entropy) in four dimensions. In this model it thus appears that 4D geometry is characterized by the lowest entanglement entropy. 

The combinatorial quantum gravity model we consider is defined by the partition function 
\begin{eqnarray}
Z &&= \sum_{\rm CS} {\rm exp} \left( -S_{EH}/\hbar \right) \ ,
\nonumber \\
S_{\rm EH} &&= - {1\over 2g} {\rm Tr}\  w^4  \ ,
\label{new1}
\end{eqnarray}
on a configuration space CS restricted to dilute random regular bipartite graphs with adjacency matrices denoted by $w$. Bipartite graphs have no odd cycles, the smallest, "elementary" loops being thus 4-cycles, squares. By "dilute" graphs we mean graphs in which two different elementary squares can share maximally one edge. This is a loop-equivalent  of a hard core requirement in a classical gas: the elementary constituents cannot overlap.
We will also focus on even connectivities of the regular graphs and denote these by $k=2d$.

Random graphs are very different from simplicial complexes, which are regular configurations that can always be associated to a geometric realization. They are ``small worlds", i.e. their diameter and average distances on the graph scale logarithmically with the number N of vertices (the volume) and they have locally a tree structure with very sparse short cycles governed by a Poisson distribution \cite{wormald} with mean $(2d-1)^l/l$ for cycles of length $l$. Because of their random character, the standard Regge formulation of discrete curvature \cite{regge} is no more applicable, a purely combinatorial version of Ricci curvature is needed. Recently, exactly such a combinatorial Ricci curvature has been proposed by Ollivier \cite{olli1}. The Ollivier curvature is very intuitive but, in general not easy to compute and work with. Fortunately, it becomes much simpler for bipartite graphs \cite{olli2}. In \cite{tru2} it was shown that 
\begin{equation}
S_{EH} = -{d^2\over g} \Big[ \sum_i \kappa (i) + {6d-3\over d} N \Big] \ ,
\label{einstein}
\end{equation}
where $\kappa(i)$ is the Ollivier combinatorial curvature scalar associated with vertex $i$ and 
\begin{equation}
\sum_i \kappa (i) = {-4\over d^2} \left[ {d(d-1)\over 2} N-N_s \right] \ ,
\label{totcursca}
\end{equation}
with $N_s\le d(d-1)N/2$ the total number of 4-cycles (squares) on the graph. Thus, apart from an irrelevant constant that we shall henceforth neglect, $S_{EH}$ is indeed a combinatorial version of the Einstein-Hilbert action. 

In \cite{tru2} it was shown in Monte Carlo simulations for finite $N$ that there are two clearly identified regimes for large $\hbar g$ and small $\hbar g$. Random graphs with logarithmic distance scaling for strong gravity (large $\hbar g$) are turned into ordered geometric graphs locally homeomorphic to $\mathbb{Z}^d$ and with power-law distance scaling $N^{1/d}$ by the condensation of squares when gravity becomes weak (small $\hbar g$). In the continuum, ordered phase the action becomes the standard Einstein-Hilbert term.The behaviour obtained is strongly suggestive of a quantum phase transition driven by the rescaled gravitational coupling $\hbar g/N^{1-2/d}$ and with order parameter $\left[2/(d(d-1))\right] N_s/N$, although present data are not sufficient to definitively conclude so. If confirmed, this quantum critical point would define gravity non-perturbatively. 

In this model geometric space emerges from the ordering of random graphs. Every classical graph, however can be viewed as the entanglement in the quantum superposition of two disconnected quantum states \cite{zan}. Consider two copies ${\cal H}_1$ and ${\cal H}_2$ of an $N$-dimensional Hilbert space ${\cal H} \simeq \mathbb{C}^N$ with basis states $ |i\rangle = (0, \dots, 1_i,\dots,0)$ for $i=1, \dots, N$. Out of these two copies let us build the pure state 
\begin{equation}
w= {1\over ||w||_F} \ \sum_{i,j=1}^N w_{ij} |i\rangle |j\rangle \ ,
\label{state}
\end{equation}
where $w$ is the adjacency matrix of the graph and  $||w||_F = \sqrt{{\rm Tr} \ w^T w}$ its Frobenius norm (here we will consider only symmetric adjacency matrices with vanishing diagonal terms, corresponding to undirected graphs with no loops). This state can be considered either as the quantum state of two $N$-level systems or the quantum state of two $N$-qubit states each constrained to the one-excitation manifold.  The important point is that the information contained in the adjacency matrix of the classical graph is equivalent to a particular entanglement pattern for the quantum state (\ref{state}). It is this beautiful equivalence, first established in \cite{zan} that permits to interpret emergent space, as encoded in the ordering of random graphs, as a quantum entanglement pattern, an approach also recently suggested in \cite{carrol}.  

Let us suppose that one of the two quantum subsystems is not observable, e.g. the one labeled by "2". The reduced density matrix for the observable quantum system "1" is then given by 
\begin{equation}
\rho \equiv \rho_1 = {\rm Tr}_2 \ |w\rangle \langle w| = {w \ w^T\over {\rm Tr} \ w^T w} \ .
\label{reddens}
\end{equation}
In the following we will focus on the quantum entanglement properties related to the graphs of the model \cite{tru2}, $2d$-regular bipartite graphs, as encoded in the von Neumann entanglement entropy $S=- {\rm Tr}\  \rho \ {\rm log}_2 \rho $. Note that these are insensitive to all re-labelings of the vertices and depend thus only on the isomorphism class of the particular graph. 

To compute the von Neumann entropy we consider first the Renyi entropies 
\begin{equation}
S_{\alpha}= {1\over 1-\alpha} {\rm log}_2 {\rm Tr} \rho^{\alpha} \ .
\label{rendef}
\end{equation}
from which the von Neumann entropy can be obtained as $\lim_{\alpha \to 1} S_{\alpha}$. Using (\ref{reddens}) we obtain
\begin{equation}
S_{\alpha}= {{\rm log}_2 {\rm Tr} \ w^{2\alpha}-\alpha \ {\rm log}_2 {\rm Tr} \ w^2 \over 1-\alpha}\ .
\label{renexp}
\end{equation}
As for all other classical graphs considered \cite{zan}, it turns out that also in the present case all entanglement entropies for integer $\alpha = k$ behave as
\begin{equation}
S_k = {\rm log}_2 N - \gamma_k + O\left( {1\over n} \right) \ ,
\label{behaviour}
\end{equation}
including the von Neumann entropy $S$, formally denoted by $k=1$ here. 
The first term simply reflects the maximum number of available qubits, whereas the second terms encodes universal properties of the particular graph class under consideration. Since they quantify the deviation from maximal entanglement in the limit $N\to \infty$ it is these terms that are universal characterizations of space in the IR limit of discrete quantum gravity models. 

Before proceeding to the limit defining the von Neumann entropy, let us compute analytically the first couple of universal Reny entanglement terms for $k = 2$ and $k =3$. To this end we use known generic formulas for the number of short cycles \cite{squares} as adapted to $2d$-regular bipartite graphs,
\begin{eqnarray} 
{\rm Tr} \ w^4 &&= 8N_s +d^2 N -2dN \ ,
\nonumber \\
{\rm Tr} \ w^6 &&= 12 N_h +(96d-48) N_s - 44 d^3N + 18 d^2 N + 4 dN \ ,
\label{traces}
\end{eqnarray}
where $N_s$ denotes the number of 4-cycles, squares and $N_h$ the number of 6-cycles, hexagons. This gives 
\begin{eqnarray}
\gamma_2 &&= {\rm log}_2 \left( {2\over d^2} {N_s\over N}  +{d-2\over 4d} \right) \ ,
\nonumber \\
\gamma_3 &&= {\rm log}_2 \sqrt{ {3\over 4d^2} {N_h\over N} + {12d-6\over d^2} {N_s \over N} - {22d^2 -9d-2\over 4d^2}} \ .
\label{gammas}
\end{eqnarray}
These expressions show that $\gamma_k$ for $k\ge2$ are geometric invariants of emergent space. They depend on the relative number of short cycles and can thus be considered quantum order parameters distinguishing the random phase with logarithmic distance scaling from the geometric lattice-like phase with power-law distance scaling. 

The same is true for the universal term in the von Neumann entropy. To obtain it we compute the limit $\alpha \to 1$ of (\ref{renexp}) and we subtract the leading term ${\rm log}_2 N$. To do so we first write $\alpha = 1+\epsilon/2$ and we then express
\begin{equation}
w^{2\alpha} = w^2 (1 + \epsilon \ {\rm ln} w) \ ,
\label{first}
\end{equation}
where we stress that the logarithm is a natural one. Therefore
\begin{eqnarray}
{\rm log}_2 {\rm Tr} \  w^{2\alpha} &&=  {\rm log}_2 \left( {\rm Tr}\ w^2 + \epsilon\  {\rm Tr} \ w^2 {\rm ln} w \right) 
\nonumber \\
&&= {\rm log}_2 {\rm Tr} \ w^2 \left( 1 + \epsilon\  {{\rm Tr} \ w^2 {\rm ln} w \over {\rm Tr} \ w^2} \right) \ ,
\label{second}
\end{eqnarray}
Inserting this into (\ref{renexp}) and taking the limit $\epsilon \to 0$ we obtain the von Neumann entropy
\begin{equation}
S= {\rm log}_2 {\rm Tr} \ w^2 - {2\over {\rm ln(2)}} {{\rm Tr} \ w^2 {\rm ln} w \over {\rm Tr} \ w^2} \ .
\label{neumann}
\end{equation}
The trace of the square of the adjacency matrix is twice the number of edges in the graph, $2dN$ for our $2d$-regular graphs with exactly $dN$ edges. 
Remembering finally that ${\rm ln}(x) = {\rm ln} (2) \ {\rm log}_2  (x)$ we obtain the final result
\begin{equation}
\gamma = \gamma_1 =  {{{\rm Tr}\  w^2  {\rm log}_2 w^2}\over {\rm Tr}\  w^2 }\ - {\rm log}_2 2d\ ,
\label{topne}
\end{equation}
which defines the topological network entanglement and plays a similar role as the universal entanglement term \cite{kit} of topologically ordered matter \cite{top}. 

The topological network entanglement $\gamma$ as a function of the connectivity parameter $d$, which corresponds to the space dimension in the geometric phase, is shown in Fig. 1. This quantity has been computed numerically as follows. First one identifies the adjacency matrix of a geometric random lattice of connectivity $2d$ on a flat $d$-dimensional torus by its equivalence with a circulant graph \cite{circulant}. The corresponding topological network entanglement $\gamma$ is then obtained by computing (\ref{topne}). Then one chooses a large value of the quantum gravity coupling $g$ in (\ref{einstein}) and one lets the initial geometric graph evolve according to a Metropolis Monte-Carlo algorithm with (\ref{einstein}) as the energy determining the acceptance or rejection of changes to the initial geometric adjacency matrix. During the evolution one can periodically measure the number of squares (4-cycles), which starts off by construction from the maximum number $N_s = d(d-1)N/2$ in the geometric phase. If the coupling $g$ has been chosen large enough, the number of squares will start to diminish until it has reached the known expectation value $(2d-1)^l/l$ (independent of $N$) of bipartite random regular graphs \cite{wormald}. At this point the evolved adjacency matrix corresponds to a bipartite random regular graph and one can compute the corresponding topological network entanglement $\gamma$ again by eq. (\ref{topne}). Repeating this process over several realizations leads to Fig. 1.

\begin{figure}
\includegraphics[width=8cm]{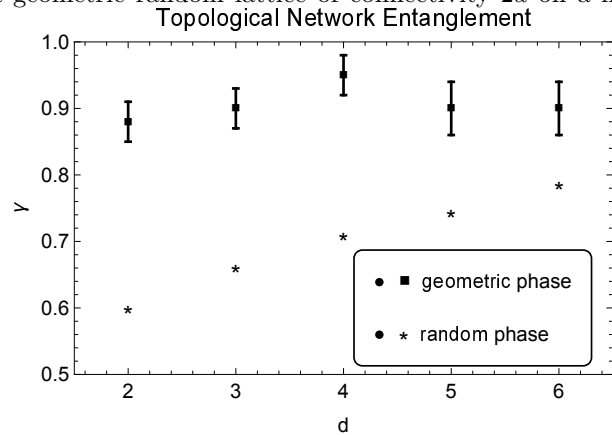}
\caption{\label{fig:Fig. 1} The topological network entanglement $\gamma$ as a function of the connectivity parameter $d$ ($k=2d$) which becomes the space dimension in the geometric phase. We have not displayed the error bars for the results in the random phase since they are about ten times smaller than their counterparts in the geometric phase and would just smear the results.}
\end{figure}

As anticipated, the emergence of geometric order increases the topological network entanglement, thereby decreasing the overall entanglement entropy. This effect is largest in four dimensions, where the entanglement entropy of the geometric phase has a local minimum.


\begin{references}
\bibitem{rams1} M. van Raamsdonk, {\it Gen. Rel. Grav.} {\bf 42} (2013) 2323. 
\bibitem{rams2} T. Faulkner, M. Guica, T. Hartmann, R. C. Myers and M. van Raamsdonk, {\it JHEP} {\bf 3} (2014) 1. 
\bibitem{rams3} N. Lashkari, M. B. McDermott and M. van Raamsdonk, {\it JHEP} {\bf 4} 1 (2014). 
\bibitem{mald} J. Maldacena, {\it JHEP} {\bf 2003}  021 (2003).
\bibitem{tru1}C. A. Trugenberger, {\it Phys. Rev.} {\bf D92} (2015) 084014, {\it Phys. Rev.} {\bf E92} (2015) 062818, {\it Phys. Rev.} {\bf E94} (2016) 052305. 
\bibitem{graphrev}For a comprehensive review see: R. Albert and L. Barabasi, {\it Rev. Mod. Phys.} {\bf 74} (2002) 47. 
\bibitem{olli1}Y. Ollivier, {\it J. Funct. Anal.} {\bf 256} (2009) 810; Y. Ollivier, {\it Adv. Stud. Pure Math.} {\bf 57} (2010) 343; Y. Linn, L. Lu and S. T. Yau, {\it Tohoku Math. J.} {\bf 63} (2011) 605; B. Loisel and P. Romon, {\it Axioms} {\bf 3} (2014) 119; J. Jost and S. Liu, {\it Discrete Comput. Geom.} {\bf 51} (2014) 300. 
\bibitem{geom}For a review see: M. Penrose, {\it Random Geometric Graphs}, Oxford University Press, Oxford (2003). 
\bibitem{tru2}C. A. Trugenberger, arXiv:1610.05934.
\bibitem{olli2}B. B. Bhattacharya and S. Mukherjee, {\it Discrete Mathematics} {\bf 338} (2015) 23. 
\bibitem{krioukov}J. Dall and M. Christensen, {\it Phys. Rev. } {\bf E66} (2002) 016121; D. Krioukov, {\it Phys. Rev. Lett.} {\bf 116} (2016) 208302. 
\bibitem{zan}S. Garneroni, P. Giorda and P. Zanardi, {\it New Journal of Physics} {\bf 14} (2012) 013011. 
\bibitem{top} For a review see: X.-G. Wen, {\it ISRN Condensed Matter Physics} {\bf 2013} (2013) 198710.
\bibitem{kit} A. Kitaev and J. Preskill, {\it Phys. Rev. Lett.} {\bf 96} (2006) 110404. 
\bibitem{wormald}P. E. O'Neil, {\it Bull. Am. Math. Soc.} {\bf 75} (1969) 1276; N.C. Wormald, {\it Surveys in Combinatorics}, LMS Lectures Note Series {\bf 267}, J. D. Lamb and D. A. Pierce eds., (1999) 239. 
\bibitem{regge}T. Regge, {\it Nuovo Cim.} {\bf 19} (1961) 558; for a review see e.g.: R. M. Williams and P. A. Tuckey, {\it Class. Quant. Gravity} {\bf 9} (1992) 1409. 
\bibitem{carrol}C. Cao, S. M. Carroll and S. Michalakis, {\it Phys. Rev.} {\bf D95} (2017) 024031. 
\bibitem{squares}F. Harary and B. Manvel, {\it Mat. Casopis Slov. Akad. Vied} {\bf 21} (1971) 55; N. Alon, R. Yuster and U. Zwick, {\it Algorithmica} {\bf 17} (1997) 209; S. N. Perepechko and A. N. Voropaev, {\it The Number of Fixed Length Cycles in an Undirected Graph. Explicit Formulas in Case of Small Length}, in {\it Mathematical Modeling and Computational Physics}, Dubna, Russia (2009). 
\bibitem{circulant}S. I. R. Costa, J. E. Strapasson, M. M. S. Alves and T. B. Carlos, {\it Linear Algebra Appl.} {\bf 432} (2010) 369; J. Marklof and A. Str\"ombergsson, {\it Combinatorica} {\bf 33} (2013) 429. 

\end{references}
\end{document}